\documentclass[12pt,twoside,letterpaper]{article}

\usepackage{graphicx}
\usepackage{fullpage}
\usepackage{xspace}
\usepackage{lineno}
\usepackage{amsmath}
\usepackage{sectsty}
\usepackage{xcolor}
\usepackage{epstopdf}

\sectionfont{\fontsize{14}{15}\selectfont}
\subsectionfont{\fontsize{12}{15}\selectfont}

\newcommand{\cls}{\ensuremath{CL_S}\xspace}
\newcommand{\pp}{\ensuremath{p\bar{p}}\xspace}
\newcommand{\cdf}{CDF\xspace}
\newcommand{\dzero}{D\O\xspace}
\newcommand{\gevcc}{\ensuremath{\mathrm{GeV}/c^2}\xspace}
\newcommand{\bb}{\ensuremath{b\bar{b}}\xspace}
\renewcommand{\ll}{\ensuremath{\ell^+\ell^-}\xspace}
\newcommand{\lv}{\ensuremath{\ell\bar{\nu}}\xspace}
\newcommand{\vv}{\ensuremath{\nu\bar{\nu}}\xspace}
\newcommand{\llbb}{\ensuremath{\ll+\bb}\xspace}
\newcommand{\lvbb}{\ensuremath{\lv+\bb}\xspace}
\newcommand{\vvbb}{\ensuremath{\vv+\bb}\xspace}
\newcommand{\ww}{\ensuremath{W^+W^-}\xspace}
\newcommand{\hbb}{\ensuremath{H\to\bb}\xspace}
\newcommand{\hww}{\ensuremath{H\to\ww}\xspace}
\renewcommand{\b}{\ensuremath{b}\xspace}
\newcommand{\vh}{\ensuremath{V\!H}\xspace}
\renewcommand{\tt}{\ensuremath{t\bar{t}}\xspace}
\newcommand{\ttH}{\ensuremath{\tt H}\xspace}
\newcommand{\ZH}{\ensuremath{Z\!H}\xspace}
\newcommand{\WH}{\ensuremath{W\!H}\xspace}

\begin{document}


\begin{center}
  \large \textbf{Standard Model Higgs boson searches at the Tevatron} \\
  \quad \\
  \normalsize Kyle J. Knoepfel \\ 
  \quad \\
  \footnotesize \textit{Fermi National Accelerator Laboratory}\\
  \footnotesize \textit{Batavia, Illinois, 60150}\\
  \footnotesize \texttt{knoepfel@fnal.gov}
\end{center}

\begin{abstract}
We give an overview of Standard Model Higgs boson studies performed at
the CDF and \dzero experiments at the Tevatron proton-antiproton
collider.  Combining the results of many individual analyses, most of
which use the full data set available, an excess with a significance
of 3.0 standard deviations with respect to the Standard Model
hypothesis is observed at a Higgs boson mass of 125~\gevcc.  At that
mass, the combined best-fit cross section is consistent with the
Standard Model prediction.  Constraints are also placed on the Higgs
boson couplings with fermions and electroweak vector bosons and are
consistent with the Standard Model predictions within the
uncertainties.
\end{abstract}

\section{Introduction}

The Higgs mechanism, first postulated in
1964~\cite{bib:englert,bib:higgs,bib:guralnik}, is thought to be
responsible for electroweak symmetry breaking in the context of the
Standard Model (SM) of particle
physics~\cite{bib:glashow,bib:weinberg,bib:salam}. In addition to
providing mass terms in the quantum field theoretic Lagrangian
density, the mechanism also yields an additional spin-zero particle.
In a formal sense, the mass of this particle, the SM Higgs boson, is
an unconstrained parameter of the theory.  However, various
electroweak observables, such as the masses of the $W$ boson and top
quark, can place indirect constraints on its mass.  Omitting direct
searches for the Higgs boson, its predicted mass is
$80^{+30}_{-23}$~\gevcc, based on a fit to measurements of many
electroweak quantities~\cite{bib:electroweak}.

In July 2012, the ATLAS and CMS experiments at the Large Hadron
Collider (LHC) reported independent observations of a particle that
was consistent with the SM Higgs boson at a mass of
125~\gevcc~\cite{bib:atlas,bib:cms}, in decent agreement with the
electroweak fit prediction.  Shortly thereafter, the CDF and \dzero
experiments at the Tevatron published a joint result showing evidence
for the Higgs boson in the \hbb search channel~\cite{bib:tevhbb},
strongly supporting the LHC discoveries.  Now that a Higgs boson has
been discovered, the best-fit cross sections and couplings can be
compared with the predicted 125-\gevcc values, both at the LHC and at
the Tevatron, providing input on whether the observed particle is
consistent with SM expectations.

This paper gives an overview of Higgs boson studies performed at the
Tevatron collider at Fermi National Accelerator Laboratory.  After
briefly introducing the Tevatron and the detectors in
Sec.~\ref{sec:tev}, we discuss the Higgs boson search channels and the
corresponding primary backgrounds in Sec.~\ref{sec:search}.  We
describe the general analysis method of searching for a Higgs boson in
Sec.~\ref{sec:analysis}, and the statistics technology required for
extracting results in Sec.~\ref{sec:stats}.  Lastly, results including
the upper limits, best-fit cross sections, and Higgs boson coupling
constraints are presented in Sec.~\ref{sec:results}.  Full details can
be found in Ref.~\cite{bib:prdTevatron} and the references therein.
For this paper, we present results only in the context of an SM
interpretation.

\section{The Tevatron and the Detectors}
\label{sec:tev}

The Tevatron was a \pp hadron collider that operated at Fermi National
Accelerator Laboratory from 1982 to 2011.  The total collision energy
in the \pp center-of-mass frame ($\sqrt{s}$) for Run II (2001 to 2011)
was 1.96 TeV.  The collision signatures were collected by two
experiments---the Collider Detector at Fermilab (\cdf) and the \dzero
experiment---which were positioned around two of the six possible \pp
interaction regions at the Tevatron.  Both experiments were
multipurpose detectors that contained various systems designed to
measure charged-particle momenta, trajectories, and energy deposits.
An important component of both detectors was the silicon sensors,
which were necessary for identifying secondary vertices displaced from
the primary interaction points, thus allowing an inference of the
presence of jets that originated from $b$ quarks, which are long-lived
compared to light-flavor quarks.

\section{Higgs Boson Search Channels}
\label{sec:search}

\begin{figure}[tbh]
\includegraphics[width=0.48\textwidth]{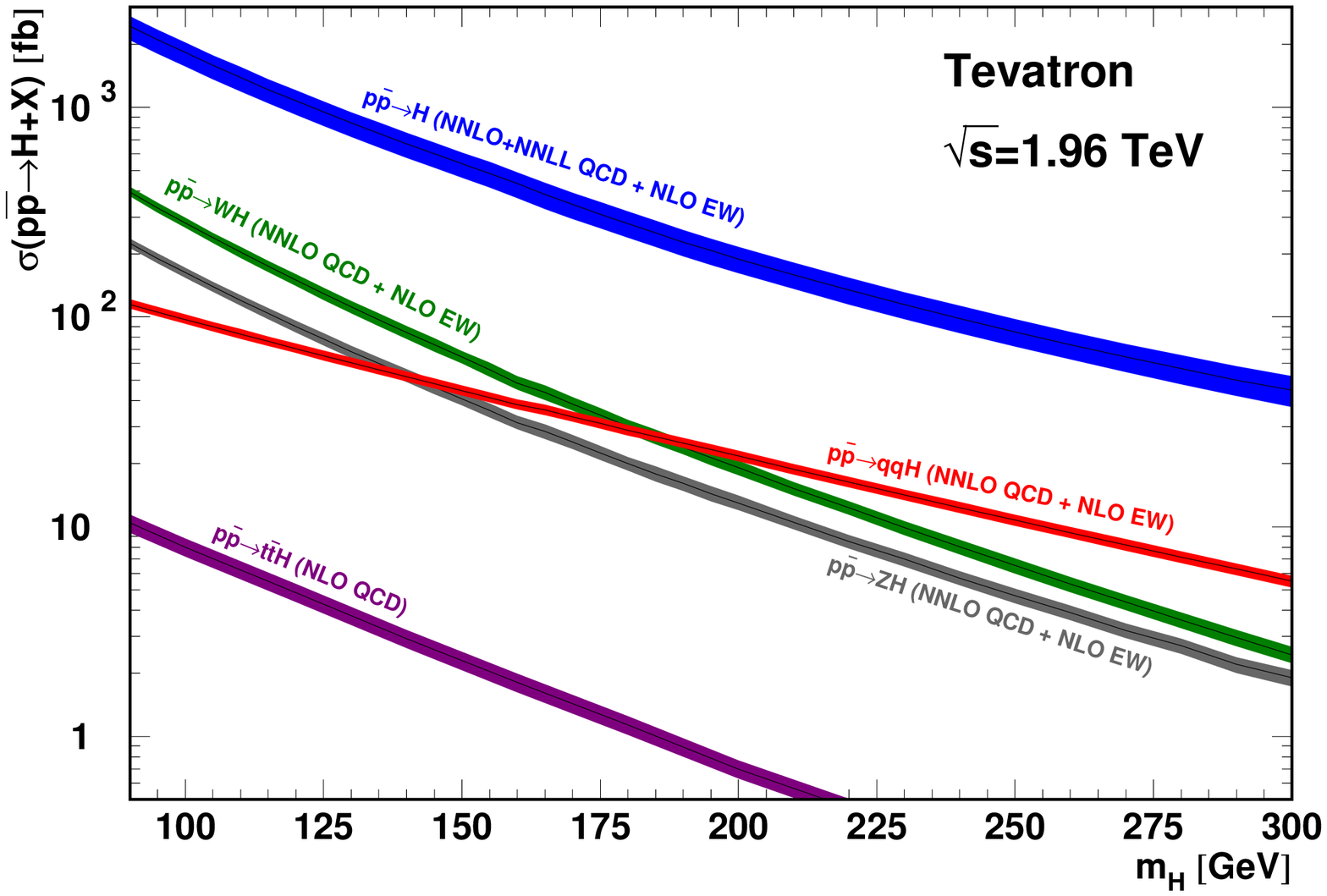}\hfill
\includegraphics[width=0.48\textwidth]{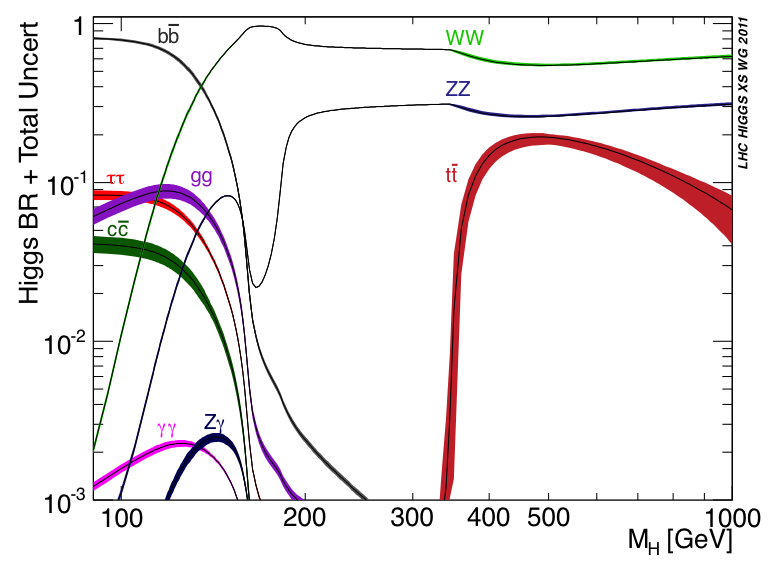}
\vspace*{8pt}
\caption{Standard Model predictions for the (left) Higgs boson
production cross section (figure from Ref. 12), and (right) the Higgs
boson branching ratio (figure from Ref. 13) as a function of the
assumed Higgs boson mass.\protect\label{fig:predictions}}
\end{figure}

As shown in the left plot of Fig.~\ref{fig:predictions}, the Higgs
boson production cross sections for \pp collisions at $\sqrt{s} =
1.96$ TeV are on the order of 0.01 to 1 pb, depending on the
production mode and the assumed Higgs boson mass in the range $100 \le
m_H \le 200$~\gevcc.  The dominant production mode is that of gluon
fusion ($gg\to H$ or $\pp\to H$), whereas the production modes where
the Higgs boson is produced in associated with an electroweak boson
$V$ (where $V = W$ or $Z$) are suppressed by roughly an order of
magnitude but benefit from triggerable signatures from the leptonic
decay of the electroweak boson.  Additional production modes include
vector-boson fusion, where the radiation of two electroweak bosons off
of the incoming \pp pair fuse to form a Higgs boson ($\pp \to q\bar{q}
H$), and the \ttH production mode, where the Higgs boson is produced
alongside a \tt pair.

The complete listing of Higgs boson searches published from the
Tevatron are listed in Ref.~\cite{bib:prdTevatron}, which includes
the references for each individual analysis and describes in some
detail the overall search strategy.  Most of the analyses use the full
data set available for the given experiment, which corresponds to
roughly $10~\mathrm{fb}^{-1}$ of integrated luminosity.  This document
will only briefly review the general methodology of searching for the
Higgs boson at the Tevatron experiments, leaving specific details to
Ref.~\cite{bib:prdTevatron} and the references therein.  Note that
in what follows, the sensitivity of an analysis typically means the
sensitivity to {\it exclude} the Higgs boson hypothesis.

\subsection{High-mass and Low-mass Searches}

Higgs boson searches at the Tevatron are classified as high-mass
(low-mass) searches if the assumed Higgs boson mass is greater than
(less than) 135~\gevcc, which corresponds to the mass where the \hbb
and \hww branching ratios are approximately equal (see right plot of
Fig.~\ref{fig:predictions}).  For high-mass searches, the \hww search
channel is the most sensitive due to the large \hww branching ratio.
The \hww final state includes two charged leptons and a significant
amount of missing transverse energy due to the neutrinos from the
$W\to\lv$ decays (where $\ell$ is an electron or muon).  For low-mass
searches, the dominant decay mode is due to \hbb, where the \bb pair
fragments and then hadronizes into a pair of jets, which (as described
in Sec.~\ref{sec:btag}) can be identified as originating from $b$
quarks.  To reduce background contributions from the strong
interaction---hereafter, \textit{QCD multijet} background---the \hbb
process is searched for in association with a leptonic decay of an
electroweak vector boson.  Low-mass searches are therefore typically
separated into three categories based on the number of reconstructed
charged leptons: $\ZH\to\llbb$, $\WH\to\lvbb$ and $\ZH\to\vvbb$
analyses, where events with misreconstructed charged leptons are
included in the acceptance of the latter two analyses.

Additional low-mass analyses are performed at the Tevatron, such as
searches for the $H\to\tau^+\tau^-$ and $H\to\gamma\gamma$ processes.
But as the sensitivities of these searches are significantly less than
those described above, we refer the interested reader to
Ref.~\cite{bib:prdTevatron} for more details.

In general, hadronic $\tau$ decays are typically not reconstructed due
to significant uncertainties in the reconstructible energy.  Rather,
for Higgs boson searches with charged leptons in the final state,
$\tau$ lepton decays are partially included from $\tau\to\lv$
processes, where the electron and muon four momenta are well
determined.

\subsection{Primary Backgrounds}

The types of backgrounds encountered in a Higgs boson analysis are
highly dependent on the search channel.  Whereas the identification of
electrons and muons is fairly robust due to the distinct signatures in
the calorimeter and muon detectors, respectively, the association of
missing transverse energy as originating from neutrinos is not
necessarily well established.  Significant variations in the
jet-energy scale (or jet-energy resolution) for QCD multijet events
can create energy imbalances that result in final state signatures
that mimic the signal process.  To suppress these backgrounds, a
typically large value of the missing transverse energy is required (50
to 60 GeV).

Other backgrounds encountered include electrons and photons that mimic
jet signatures, and various irreducible backgrounds, such as Drell-Yan
processes ($\pp\to\ll+X$), and the SM diboson contributions $WW$, $WZ$
and $ZZ$, whose final states are largely indistinguishable from those
of the associated Higgs boson production modes.  Many analyses have
custom event selection requirements aimed at suppressing these
backgrounds to the extent possible.

\section{Analysis Methods}
\label{sec:analysis}

The analysis procedures for each Higgs boson search are similar: after
imposing a set of event selection criteria on the triggered data, the
remaining events are generally separated into those that fall within a
signal region, where the measurement is performed, and those that
populate various control regions, which are used to validate the
background model and analysis methods.  In the last few years of Higgs
boson analyses at the Tevatron, both experiments have increased signal
acceptance by relaxing the lepton-identification and kinematic
criteria.  The sensitivity to excluding the Higgs boson is further
improved by optimizing \b-jet identification (for low-mass analyses)
and by finding ways to separate the signal and background processes in
the final histogram used for extracting results, as discussed below.

\subsection{\boldmath \b-jet identification}
\label{sec:btag}

The ability to identify a jet as originating from the fragmentation of
a $b$ quark is central to searching for the Higgs boson in the \hbb
channel.  This is achieved by taking advantage of the longer lifetime
of the $b$ quark compared to that of a light-flavor quark.  With an
average lifetime on the order of 1.5 ps, the $b$ quark can travel on
the order of a few mm in the transverse plane with respect to the
primary interaction vertex before it decays.  By using the fine
granularity of the silicon detector, the decay vertex of the $b$ quark
can be inferred with high confidence, thus allowing a potential
matching between the $b$ quark and the corresponding jet as detected
by the calorimeter.  The goal is to maximize the $b$-jet
identification (or $b$-tag) efficiency, while minimizing the
false-positive (or mistag) rate.

The final $b$-tagging algorithm used by CDF was specifically optimized
for Higgs boson searches~\cite{bib:hobit}. It was a neural network that
incorporated inputs from several $b$-tagging algorithms that preceded
it, leveraging the correlations between them to further increase
$b$-tagging efficiency.  The \dzero approach was similar, but used a
boosted decision tree algorithm instead.  For both experiments, a
\b-tag efficiency of better than 50\% could be obtained for a
corresponding mistag rate of 1\%.

\subsection{Multivariate techniques}

To maximize the sensitivity to excluding the Higgs boson hypothesis, a
discriminant function is constructed that seeks to separate the Higgs
boson signal from the backgrounds that obscure it.  Each event is then
assigned a discriminant value and then binned into a histogram that is
used to extract results.  Various methods are used to construct the
discriminant function.  The Tevatron analyses typically use
multivariate machine-learned algorithms such as neural networks or
boosted decision trees, which are trained using Monte Carlo simulated
events that are classified as signal or background.  The algorithm
then ``learns" how to distinguish between signal and background events
based on the variables that serve as inputs to the discriminant
function.  An optimal algorithm will result in signal and background
distributions of the discriminant value that are maximally separated.
This can be approximately achieved by using input variables that
exhibit significantly different shapes for signal and background
processes.

Many of the Tevatron analyses use several discriminants in a cascade
manner.  To illustrate, one discriminant may be designed to
(\textit{e.g.}) separate the signal from QCD multijet background; a
threshold criterion is then required of the discriminant value for the
event.  For the surviving events, another discriminant is designed to
separate the signal from the remaining backgrounds.  The histogram
corresponding to this final discriminant is then used for obtaining
results.  Examples of analyses that use this cascade method are the
CDF $\ZH\to\llbb$~\cite{bib:cdfllbb} and the \dzero
$\ZH\to\vvbb$~\cite{bib:d0metbb} searches.

Note that the discriminant used in obtaining results is not always the
binned histogram of a value that only has an abstract meaning (as in
the case of machine-learned nonlinear algorithms).  Although a
nonlinear algorithm can typically take into account nontrivial
correlations between kinematic variables, usually improving the
analysis sensitivity, a simple physical quantity can sometimes be
nearly as effective.  For example, the invariant diphoton mass for the
$H\to\gamma\gamma$ searches~\cite{bib:diphotonCDF,bib:diphotonDzero} is
used to extract results instead of a multivariate algorithm.

Sections~\ref{sec:stats} and~\ref{sec:results} describe how the
discriminants are used to obtain results.

\section{Statistical Methods}
\label{sec:stats}

For each individual analysis, a likelihood is constructed
\begin{equation}\label{eqn:like}
L = \prod_j \frac{\mu_j^{n_j}}{n_j!}e^{-\mu_j} 
\end{equation}
which is the product of Poisson probabilities for each of the bins in
the discriminant used to obtain results.  The observed and expected
number of events for each bin are represented by $n_j$ and $\mu_j$,
respectively.  In the absence of a Higgs boson signal, $\mu_j$ is
simply equal to the expected background $b_j$.  When attempting to
exclude a Higgs boson signal, $\mu_j$ is allowed to include a signal
component $\mu_j = Rs_j + b_j$ where the expected SM Higgs boson
signal $s_j$ is scaled by $R$, which assesses the sensitivity in the
data to a Higgs boson signal.

The presence of systematic uncertainties degrades the sensitivity to
excluding a Higgs boson and therefore must be taken into account.
This is achieved by introducing a set of nuisance parameters
$\boldsymbol{\theta}$ that are expected to follow Gaussian
distributions centered at the central value of a given systematic
correction, with a root-mean-square width equal to the one-standard
deviation uncertainty of the systematic effect.  A given nuisance
parameter $\theta_k$ is thus represented by a Gaussian
prior-probability density $\pi_k(\theta_k)$.  In principle, the
nuisance parameters can be correlated, in which case the appropriate
prior-probability density is a multivariate density
$\pi(\boldsymbol{\theta})$.  The likelihood is thus modified to be:
\begin{equation}\label{eqn:like2}
L = \prod_j \frac{\mu_j^{n_j}}{n_j!}e^{-\mu_j} \times \pi(\boldsymbol{\theta}) = \prod_j \frac{\mu_j^{n_j}}{n_j!}e^{-\mu_j} \prod_k \pi_k (\theta_k)
\end{equation}
where the second equality applies if $\boldsymbol{\theta}$ represents
a set of $k$ independent nuisance parameters
$\left\{\theta_k\right\}$.  As the values of $\mu_j$ are allowed to be
influenced by the nuisance parameters, the Gaussian prior densities
$\pi_k$ are truncated at zero to ensure non-negative event yields.

When combining various analysis channels an overall likelihood is
constructed, built from the individual likelihoods as given by
Eq.~(\ref{eqn:like}):
\begin{equation}\label{eqn:overlike2}
\mathcal{L} = \prod_i\prod_j \frac{\mu_{ij}^{n_{ij}}}{n_{ij}!}e^{-\mu_{ij}} \times \pi(\boldsymbol{\Theta})
\end{equation}
where an extra index $i$ has been included to account for each of the
analyses that contribute to the combination, and $\boldsymbol{\Theta}$
is the corresponding superset of all nuisance parameters.
Measurements are extracted from the data by maximizing the likelihood
in the case of the modified frequentist technique used by the \dzero
experiment, or by maximizing the posterior probability density for the
Bayesian technique used by CDF.  Both methods are described below, and
in what follows, the null (test) hypothesis refers to the SM scenario
where the expected number of events excludes (includes) Higgs boson
contributions.

\subsection{\boldmath \cls method}
\label{sec:cls}

The \cls method~\cite{bib:cls1,bib:cls2} is the chosen likelihood
maximization technique by the \dzero experiment.  It is frequentist in
the sense that no prior assumption is made on the signal strength $R$.
A log-likelihood ratio $Q$ is formed as the test statistic, where
\begin{equation}
Q(R) = -2\ln\frac{\mathcal{L}\left(\mathbf{n} | R \right)}{\mathcal{L}\left(\mathbf{n} | 0 \right)} \ .
\end{equation}
The denominator corresponds to the likelihood assuming $R$ is null,
and the numerator allows for $R$ to be positive.  The binned event
yields $\mathbf{n}$ are those that are either observed (as in data) or
are produced from pseudo-experiments, as described below.

For a given value of $R$, pseudo-experiments are produced by drawing
random variates for $\mathbf{n}$ from both the null and test
hypotheses, and then forming distributions in the variable $Q$.
According to the Neyman-Pearson lemma~\cite{bib:neymanpearson}, this
test statistic provides maximum discrimination power between the null
hypothesis $H_0$ and the test hypothesis $H_1$.  From each of these
hypotheses a $p$-value is calculated that assesses the probability
that a result as or more extreme than that observed could be produced
under the given hypothesis.  For the null hypothesis, the $p$-value is
termed $p_0$, whereas $p_1$ is used for the test hypothesis.  The
frequentist approach is to exclude the test hypothesis if $p_1$ is
less than 5\%.  It is conceivable, however, that a downward
fluctuation of the data could exclude not only the test hypothesis but
also the null hypothesis.  To avoid this situation, the \cls value is
constructed, where the standard frequentist result $p_1$ is multiplied
by $(1-p_0)^{-1}$.  This has the benefit of ensuring conservative
results, but it also effectively reduces the sensitivity to excluding
the test hypothesis.  To obtain confidence levels, the value of $R$ is
varied until the \cls quantity is less than the value corresponding to
the desired exclusion region.

\subsection{Bayesian method}

In contrast to \dzero, the CDF experiment adopts a Bayesian technique,
which requires introducing a prior probability density $\pi(R)$ that
parameterizes the ``degree of belief" the signal strength $R$ is
expected to follow.  The relevant function is therefore a multivariate
posterior probability density:
\begin{equation}
p (L,R) =  \mathcal{C}\times L(\boldsymbol{\theta}) \times \pi(R) 
\end{equation}
where $\mathcal{C}$ is a normalization factor, $L$ is defined by
Eq.~(\ref{eqn:like}), and the assumed prior density $\pi(R)$ is a
non-negative uniform distribution.  For clarity, the dependence of $L$
on $\boldsymbol{\theta}$ has been explicitly indicated.

In contrast to maximizing the likelihood $L$, the CDF approach is to
marginalize (integrate) over each of the nuisance parameters and
extract results based on the marginalized posterior probability
density function:
\begin{equation}
p (R) =  \mathcal{C} \pi(R) \int L(\boldsymbol{\theta}) d\boldsymbol{\theta},
\end{equation}
where the only variable not integrated over is the signal strength
$R$.  How this posterior probability density is used is discussed in
Sec.~\ref{sec:results}.

\section{Results}
\label{sec:results}

As mentioned earlier, the \cls and Bayesian marginalization techniques
are favored by \dzero and CDF, respectively, in extracting results on
Higgs boson production.  In practice, the agreement of the results
from both methods is better than 5\% with respect to each other.  An
{\it a priori} decision was made to quote Bayesian results in
Ref.~\cite{bib:prdTevatron}, and that decision is retained here.

\subsection{Upper limits on Higgs boson production}
\label{sec:limits}

To determine the sensitivity a given analysis has to the Higgs boson,
many pseudo-experiments are produced that estimate the values of the
signal strength one would expect to obtain if no Higgs boson were
present (the null hypothesis).  For each pseudo-experiment, the
$p_0(R)$ density is integrated from a lower bound of zero to $R_{95}$,
which corresponds to the upper value of the signal strength region
that contains 95\% of the posterior density.  As a Bayesian technique
is adopted the limit corresponds to a 95\% credibility level (C.L.)
upper limit.  The distribution of $R_{95}$ values is then used to
determine the median-expected sensitivity to the Higgs boson, and the
1- and 2-standard deviation uncertainties on its prediction.  The
value of $R_{95}$ is then determined for data, which corresponds to
the observed limit.

\begin{figure}[tbh]
\centerline{\includegraphics[width=0.8\textwidth]{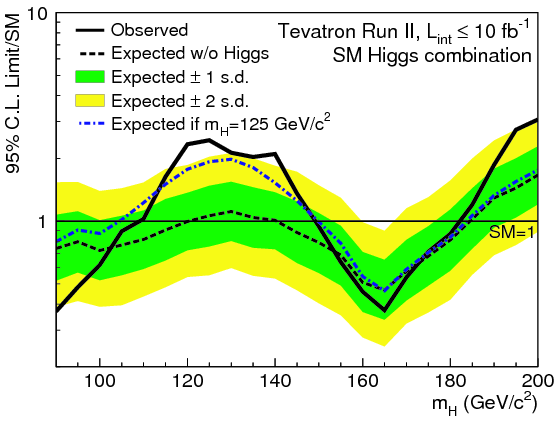}}
\vspace*{8pt}
\caption{95\% credibility level limits on SM Higgs boson production
using Tevatron data (figure from Ref. 11).\protect\label{fig:limits}}
\end{figure}

The limit-setting procedure is performed for various Higgs boson
masses, depending on the analysis in question.  For the Tevatron
combination, the union of analyses spans a Higgs boson mass range of
$90 \le m_H \le 200$ GeV$/c^2$.  The Tevatron upper limit combination
on Higgs boson production is shown in Fig.~\ref{fig:limits}.  The
vertical axis represents the value of $R_{95}$, which is in units of
the SM prediction.  In the mass range $115 \le m_H \le 140$~\gevcc,
the observed data limits are in significant disagreement with the
predictions, which assume the null hypothesis.  Also plotted is a
curve of what one would expect the limit to look like if a SM Higgs
boson of mass 125 GeV$/c^2$, produced at the SM rate, were present in
the data.  As can be seen, the SM hypothesis is favored over the null
hypothesis in the mass range from roughly 115 to 140 GeV$/c^2$.  As
the primary sensitivity in this mass range results from the $\hbb$
analyses, an excess that is wide is expected if the SM Higgs boson is
present, due to significant energy resolution effects of the $b$ jets.

\subsection{Significance of the excess}

The significance of the excess can be determined by estimating the
$p$-value of the observed data under the null hypothesis.  Before
discovery of the Higgs boson by the LHC collaborations, this
calculation was performed at each Higgs boson mass under
consideration, which also entailed introducing a trials factor that
accounted for the look-elsewhere effect.  Now that the Higgs boson is
known to reside at roughly 125 GeV$/c^2$, the Tevatron experiments
quote significances just at that mass.  For a Higgs boson mass of 125
GeV$/c^2$, the Tevatron combination $p$-value corresponds to a
significance of 3.0 Gaussian standard deviations, where the expected
excess given the SM prediction is 1.9 standard deviations.

\subsection{Best-fit cross sections}

\begin{figure}[tbh]
  \includegraphics[height=0.30\textheight]{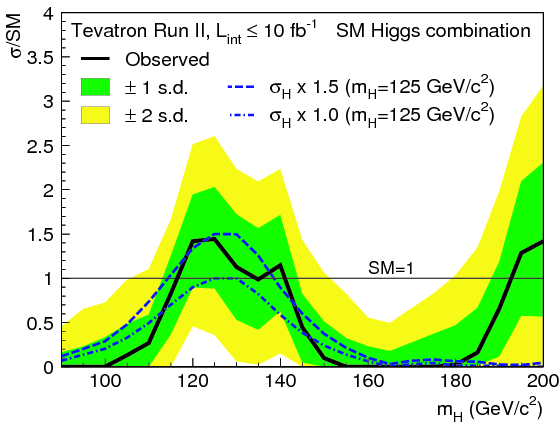} \hfill
  \includegraphics[width =0.44\textwidth ]{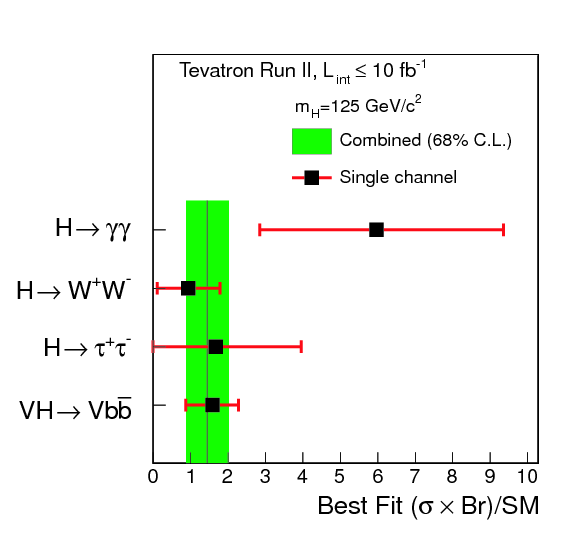}
  \caption{Best-fit cross sections using Tevatron data, assuming the SM
    Higgs boson hypothesis for (left) all Higgs boson masses considered,
    and for (right) a mass of 125 GeV$/c^2$, separated according to search
    channel (figures from Ref. 11).\protect\label{fig:xsecs}}
\end{figure}

Instead of quoting the 95\% C.L. upper limit on the null hypothesis,
the posterior distribution $p_1(R)$ under the test hypothesis can be
used to quote a ``best-fit" result, which corresponds to the signal
strength that gives the largest posterior probability density.  The
uncertainty on the best-fit result is given as the shortest $R$
interval that contains 68\% of the posterior density.  The left plot
of Fig.~\ref{fig:xsecs} shows the best-fit cross sections, and their
1- and 2-standard deviation uncertainties, across the full Higgs boson
mass range considered in Sec.~\ref{sec:limits}.  The best-fit results
are consistent with the SM prediction within uncertainties in the
Higgs boson mass range of roughly 115 to 140 GeV$/c^2$.  Also plotted
are the expected best-fit results if the Higgs boson were present at a
mass of 125 GeV$/c^2$ at the SM production rate, and also at a 50\%
greater rate, which is more in agreement with what is observed.

For a Higgs boson mass of 125 GeV$/c^2$, the best-fit cross sections
and their uncertainties are separately calculated for the
$H\to\gamma\gamma$, $\hww$, $H\to\tau^+\tau^-$, and $\hbb$ search
channels, as shown on the right plot of Fig.~\ref{fig:xsecs}.  All of
the best-fit results from the four search channels presented in
Fig.~\ref{fig:xsecs} are consistent with the SM predictions within
uncertainties.

\subsection{Constraining Higgs boson couplings}

Up until this point, two hypotheses have been considered.  Either the
null hypothesis is assumed, where one places upper limits on Higgs
boson production, or the signal hypothesis is assumed and best-fit
cross sections are extracted given the SM prediction.  One can also
test for the presence of beyond-the-SM physics by assuming the
presence of the Higgs boson, but allowing the various Higgs boson
couplings to vary.  To effect this test, $\kappa_f$, $\kappa_W$ and
$\kappa_Z$ scale factors are introduced as coefficients in front of
the $H\!f\!f$, $HWW$, and $HZZ$ couplings.  For cases when custodial
symmetry is assumed, we use $\kappa_V \equiv \kappa_W = \kappa_Z$.
The scale factors are normalized in such a way that the SM Higgs boson
couplings are recovered when $\kappa_f = \kappa_W = \kappa_Z = 1$.

When the $\kappa$ scale factors are introduced, the expressions for
Higgs boson production and decay are modified as indicated in
Table~\ref{tab:kappas}.  As the primary sensitivity to the Higgs boson
at the Tevatron is in its $\hbb$ searches, the expressions include a
cross-term from $\left(\kappa_f\kappa_V\right)^2$, where $\kappa_V$
originates from the associated production mode and $\kappa_f$ enters
from the diquark decay of the Higgs boson.  In contrast, some of the
searches that are less sensitive to Higgs boson production provide
clean avenues to constraining the individual Higgs boson couplings.
For example, the $\ttH$ search does not include any $HVV$ couplings
and is thus dependent on only $\kappa_f$ in the numerator; and the
$\vh \to V+\ww$ search does not include any $H\!f\!f$ couplings, thus
giving dependence on only $\kappa_V$ in the numerator.

\begin{table}[b]
\caption{Modification of production and decay expressions from introducing
$\kappa$ scale factors, including an approximate modification for the
total Higgs boson decay width, which appears as a denominator.}
\centering
\begin{tabular}{lcr} 
  \hline\hline
  Process &  \multicolumn{2}{c}{$\sigma\times\mathcal{B}_{H}$ (w.r.t. SM)} \\ 
  \hline
  $\vh \to V+\bb$ & $\propto$ &  $\left(\kappa_f\kappa_V\right)^2/\left(\frac{3}{4}\kappa_f^2+\frac{1}{4}\kappa_V^2\right)$ \\
  $\ttH \to \tt + \bb$ & $\propto$ & $\left(\kappa_f^2\right)^2/\left(\frac{3}{4}\kappa_f^2+\frac{1}{4}\kappa_V^2\right)$ \\
  $\vh \to V+\ww$ & $\propto$ & $\left(\kappa_V^2\right)^2/\left(\frac{3}{4}\kappa_f^2+\frac{1}{4}\kappa_V^2\right)$ \\
  \hline\hline
\end{tabular}
\label{tab:kappas}
\end{table}

\begin{figure}[tbh]
  \includegraphics[width=0.48\textwidth]{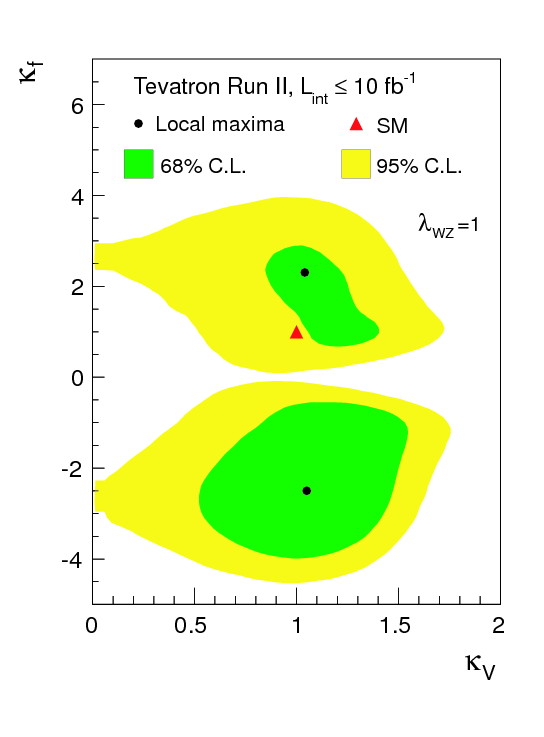}\hfill
  \includegraphics[width=0.48\textwidth]{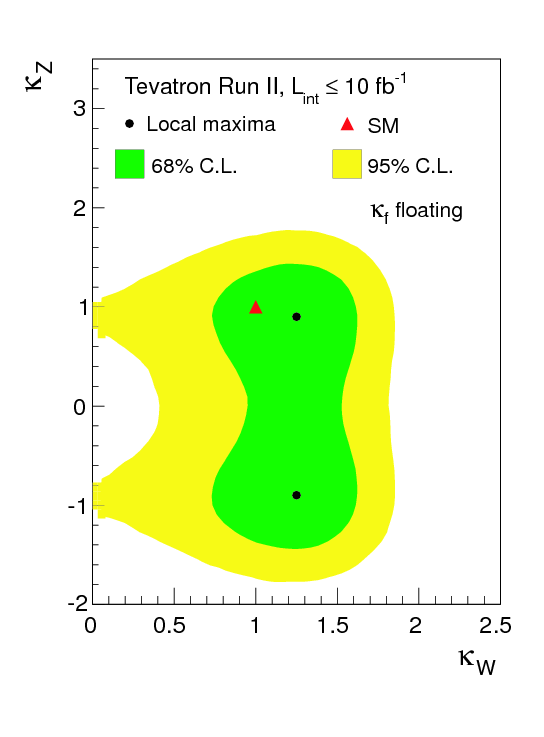}
  \caption{Two-dimensional posterior probability density functions using
    Tevatron data testing for the couplings using Tevatron data (figures
    from Ref. 11). \protect\label{fig:kappas}}
\end{figure}

The left and right plots of Fig.~\ref{fig:kappas} show the
two-dimensional posterior probability density functions of $\kappa_f$
vs. $\kappa_V$ and $\kappa_Z$ vs. $\kappa_W$, respectively.  As can be
seen, the observed results are consistent with the SM predictions at
(left) just over one standard deviation, and (right) to well within
the one standard-deviation uncertainty.

\section{Conclusions}

The CDF and \dzero Tevatron experiments have performed searches for
the SM Higgs boson across an assumed mass range of $90 \le m_H \le
200$~\gevcc.  An excess with a significance of 3.0 standard deviations
with respect to the SM hypothesis is observed at a Higgs boson mass of
125~\gevcc.  At that mass, the combined best-fit cross section is
consistent with the SM prediction.  The Tevatron experiments further
constrain the Higgs boson couplings by introducing multiplicative
corrections into the cross-section times branching ratio expressions.
The data at the Tevatron favor the SM Higgs boson couplings within the
uncertainties.

\section*{Acknowledgments}

The author would like to thank the members of the Tevatron New
Phenomena and Higgs Working Group for helpful comments and
clarifications.  The author also acknowledges the support of the CDF
department at Fermilab, and he thanks the editors of Modern Physics
Letters A for the invitation to submit this article. \medskip

\noindent This review contains copyrighted material: 
\begin{itemize}
\item Fig.~\ref{fig:predictions}(left) was reprinted with permission
from Ref.~\cite{bib:pdg}. Copyright 2012 the American Physical
Society.
\item Fig.~\ref{fig:predictions}(right) was reprinted from the Open
Access report Ref.~\cite{bib:lhcHiggsXs} under the Creative Commons
Attribution 3.0 license.  Copyright 2013 CERN.
\item Figs.~\ref{fig:limits},~\ref{fig:xsecs}, and~\ref{fig:kappas}
were reprinted with permission from
Ref.~\cite{bib:prdTevatron}. Copyright 2013 the American Physical
Society.
\end{itemize}

\end{document}